\documentclass[prb]{revtex4}

\usepackage{graphicx}% Include figure files
\usepackage{dcolumn}% Align table columns on decimal point
\usepackage{bm}% bold math

\begin{document}
\title{
Kinetic energy of He atoms in liquid $^4$He-$^3$He   mixtures}
\author{R. Senesi}
\affiliation{ Istituto Nazionale per la Fisica della Materia\\
UdR Roma Tor Vergata\\ via della Ricerca Scientifica 1, 00133
Roma, Italy } \email{roberto.senesi@roma2.infn.it}

\author{ C. Andreani}
\affiliation{ Dipartimento di Fisica\\
Universit\`{a} degli Studi di Roma Tor Vergata\\ and\\ Istituto
Nazionale per la Fisica della Materia\\ via della Ricerca
Scientifica 1, 00133 Roma, Italy }

\author{A. L. Fielding}
\affiliation{ Joint Department of Physics\\
 Institute of Cancer Research and Royal Marsden Hospital\\
 Downs Road Sutton Surrey SM2 5PT, United Kingdom}

\author{J. Mayers}
\affiliation{ Rutherford Appleton Laboratory, ISIS Facility\\
Chilton, Didcot, Oxon Ox11 0QX, United Kingdom}

\author{W. G. Stirling}

\affiliation{ European Synchrotron Radiation Facility\\
 B.P. 220, F-38043, Grenoble, Cedex France\\ and\\
 Department of Physics, University of Liverpool,\\
 Liverpool L69 3BX, United Kingdom}

\date{\today}

\begin{abstract}
 Deep Inelastic Neutron Scattering measurements on liquid
$^3$He-$^4$He mixtures in the normal phase have been performed on
the VESUVIO spectrometer at the ISIS pulsed neutron source at
exchanged wavevectors of about $q \simeq 120.0$ \AA$^{-1}$. The
neutron Compton profiles J(y) of the mixtures  were measured along
the $T=1.96$ K isotherm for $^3$He concentrations, $x$, ranging
from $0.1$  to $1.0$ at saturated vapour pressures. Values of
kinetic energies, $\langle T \rangle$, of $^3$He and $^4$He atoms
as a function of $x$, $\langle T \rangle(x)$, were extracted
 from the second moment
of J(y). The present determinations of $\langle T \rangle(x)$
confirm previous experimental findings for both isotopes and, in
the case of $^3$He, a substantial disagreement with theory is
found. In particular $\langle T \rangle(x)$ for the $^3$He atoms
is found to be independent of concentration yielding a value
$\langle T \rangle_3(x=0.1)\simeq 12 K $, much lower than the
value suggested by the most recent theoretical estimates of
approximately 19 K .

\end{abstract}


\maketitle

\section{Introduction}
 \indent The microscopic static and dynamic properties of
liquid $^4$He-$^3$He mixtures are characterized by the interplay between the
Fermi ($^3$He) and Bose ($^4$He) statistics, the interatomic interaction, and
the quantum-mechanical zero-point motion\cite{krotscheck,Dobbs}.  Moreover,
the Pauli exclusion principle strongly influences the stability of the
mixture\cite{krotscheck}. Dilute solutions of $^3$He atoms in liquid $^4$He
form a prototype quantum liquid as an example of an interacting boson-fermion
mixture. Indeed the presence of $^3$He affects the condensate fraction $n_0$,
the superfluid fraction ${\rho_s}/ {\rho_4}$ of $^4$He, the  individual
momentum distributions, n(p),and the single-atom mean kinetic energy,
$\langle T \rangle$, of the two isotopes. In recent years, considerable
efforts have been addressed to the understanding of microscopic static and
dynamical properties in helium mixtures from both the experimental and the
theoretical points of view \cite {Dobbs}. Experimental Deep Inelastic Neutron
Scattering
 (DINS) results have revealed significant and interesting discrepancies between
theory and experiment as far as determination of the condensate fraction
$n_0$ in the superfluid phase, mean kinetic energy $\langle T \rangle_3(x)$
of the lighter isotope, and momentum distributions  are concerned
\cite{az3he4he,wang}. Here, $x$ is the concentration of $^3$He in the
mixture. We stress that a substantial agreement between theories exists for
the values of $n_0$ and $\langle T \rangle_3(x)$ for low concentration
mixtures \cite{bor97,mazzanti,boninsegni,lee,ceperleymix}. This discrepancies
have to be compared with the remarkable agreement found between theory and
experiments for pure high density liquid and solid
$^3$He\cite{Dobbs,senesi,casulleras,moroni}
 and pure fluid and solid $^4$He\cite{glyde,glydecond,zoppi,ceperley}, respectively.

The single-atom mean kinetic energies $\langle T \rangle(x)$ reflect the
localization of the two isotopes in the mixture and are influenced by the
mixture concentration\cite{az3he4he,wang,bor99,momentum}. An important
conclusion of DINS measurements in the concentration range $0.0 \leq x \leq
1.0$ \cite{az3he4he,wang}, is that $\langle T \rangle_3(x)$ is essentially
independent of $x$, indicating a local environment of the $^3$He atoms in the
mixtures similar to that of pure liquid $^3$He. Ground state simulation
techniques provide an insight into the local environment of
 $^3$He and $^4$He in liquid He mixtures and pure He liquids, allowing the evaluation of both partial
 radial distribution functions and single particle mean kinetic energies.
 In particular, simulation results for partial radial distribution functions,
 $g_{\alpha,\beta}(r)$,
  in low concentration mixtures ($x \leq 0.1$),
 show two distinct features. The radial distribution function $g_{3,4}(r)$ and local density
 profile, $\rho_{3,4}(r)$, are very similar to $g_{4,4}(r)$,
 $\rho_{4,4}(r)$ and to the pure $^4$He radial distribution function
  \cite{ceperleymix,fabrocini}. The second feature is that $g_{3,4}(r)$ and
 $\rho_{3,4}(r)$ are
 markedly different from $g_{3,3}(r)$, $\rho_{3,3}(r)$ and from pure
 $^3$He radial distribution function.
 These findings support a picture where the $^3$He atoms experience a greater localization
 in the mixture with respect to pure $^3$He, while the $^4$He atoms show a microscopic structure
 similar to pure liquid $^4$He. The first feature accounts for the increased
 $\langle T \rangle_3$
 values with respect to pure liquid $^3$He \cite{ceperleymix,bor97,mazzanti,boninsegni,bor99}.
  In the case of $^4$He, $\langle T \rangle_4$ is on the contrary similar to
  the pure liquid value for $x \rightarrow 0$, and a decrease of $\langle T \rangle_4$ with
   increasing concentration is found, in agreement with experimental DINS results.
  It has to be stressed that the similarity of  $g_{3,4}(r)$ and $g_{4,4}(r)$ does not necessarily imply
  similar values of $\langle T \rangle_3$ and $\langle T \rangle_4$ in the mixture.
Moreover, since the atomic density $n$ in the mixtures is always larger than
the atomic density of pure liquid $^3$He, DINS results show that $\langle T
\rangle_3$ is also independent of $n$; this, again, is in contrast with the
widely-assessed density-dependence of mean kinetic energy of all quantum
fluids and solids. As far as the experimental values of $\langle T \rangle_4
(x)$ in the $0.0 \leq x \leq 0.4$ range are concerned, these were found to be
in agreement with microscopic calculations, resulting in a decrease of the
kinetic energy of the $^4$He atoms with increasing concentration.

These findings motivated the present measurements, which were
performed over a wider concentration range, i.e. $x= 0.00, 0.10,
0.35, 0.65, 0.90,1.00$, and with an increased statistical accuracy
than previous DINS experiments \cite{az3he4he,wang}. At present
DINS is the only experimental technique which allows direct access
to single-particle dynamical properties, such as the momentum
distribution, $n(\vec{p})$, and mean kinetic energy $\langle T
\rangle$ \cite{evans}. Experimentally this is achieved by
exploiting the large values of wavevector and energy transfers
involved in neutron scattering with epithermal neutrons
\cite{momentum}. The scattering process is well described within
the framework of the {\it Impulse Approximation} (IA). In the IA
the dynamical structure factor $S(\vec{q},\omega )$, which
determines the scattered intensity, is given by:
\begin{equation}
S_{IA}(\vec{q},\omega )=\int n(\vec{p})~\delta \left( \omega -\frac{\hbar
q^{2}}{2M}-\frac{\vec{q}\cdot \vec{p}}{M}\right) d\vec{p},
\end{equation}
where $M$ is the atomic mass of the struck nucleus. The scaling properties of
the  scattering law can be
expressed in terms of a scaling function: $J(y,\hat{q})=\frac{\hbar q}{M}$ $%
S_{IA}(\vec{q},\omega )$, where $y=\frac{M}{\hbar q}\left( \omega -\frac{%
\hbar q^{2}}{2M}\right) $ is the West scaling variable
\cite{watson,momentum}. The function $J(y,\hat{q})$, often
referred to as the {\it Neutron Compton Profile} (NCP) or
longitudinal momentum distribution \cite{watson,momentum},
represents the probability density distribution of $y$, the atomic
momentum component along the direction of momentum transfer
$\hat{q}$. In the present case the dependence on the direction of
the momentum transfer $\hat{q}$ will be omitted given the absence
of preferred orientations in the liquid samples. The values for
$\left\langle T\right\rangle $\ are obtained by exploiting the
second moment sum rule for $J(y)$ \cite{glyde,watson}:
\begin{equation}
\int_{-\infty }^{\infty }y^{2}J(y)~dy=\sigma _{y}^{2}=\frac{2M}{%
3\hbar ^{2}}\langle T \rangle
\end{equation}
where $\sigma _{y}$ is the standard deviation of $J(y)$.  DINS
spectra from a liquid $^3$He-$^4$He mixture will be then composed
of two distinct contributions, one corresponding to the $^4$He
NCP, $J(y_4)$, and, in a different region of the DINS spectra, one
to the $^3$He NCP, $J(y_3)$. The $J(y_{3,4})$ functions can be
separately analyzed and from their lineshape properties the
momentum distributions and mean kinetic energies can be
determined.

\section{Experiment}

The DINS measurements were carried out on the VESUVIO instrument,
an inverse-geometry spectrometer operating at the ISIS pulsed
neutron source (Chilton, Didcot-UK)\cite{vesuvio}. On this
instrument the NCP spectrum is reconstructed using the filter
difference technique which consists of measuring the time of
flight of the neutrons scattered by the sample; the final energy
is selected by a resonant foil analyzer located between sample and
detectors \cite{evans}. For the present experiment the $4.908$ eV
resonance of a $^{197}$Au foil filter was chosen. The scattered
neutrons were detected by 32 glass scintillators ($^6$Li-enriched
fixed-angle elements) placed over an angular range $115^{o}<$
$2\theta <144^{o}$,  yielding average wavevector transfers of $q
\simeq 128$\AA $^{-1}$ and $q \simeq 116$\AA $^{-1}$ for $^3$He
and $^4$He respectively. These large values of wavevector
transfers ensured that deviations from the IA, generally described
in terms of the {\it Final State Effects} (FSE) \cite{glyde} were
negligible and did not affect significantly the recoil peak shapes
\cite{rinat,senesi}. The corresponding average energy transfers
accessed were $\hbar\omega \simeq 11 eV$ and $\hbar\omega \simeq 7
eV$ for $^3$He and $^4$He respectively. For highly absorbing
$^3$He this energy yields a favourable ratio between the
absorption and scattering cross section of about $30$
\cite{senesi}.

The experiment has been performed along the $1.96$ K isotherm.
Known amounts of gaseous $^3$He and $^4$He were first mixed in a
reservoir at $T=293 K$. Six mixtures were prepared, for different
$^3$He concentrations, namely  $x= 1.00,0.90, 0.65, 0.35, 0.10,
0.00$. The mixtures were then allowed to condense into the sample
cell to the homogeneous liquid phase at T= $1.96$ K. Special
attention was paid to ensure that the liquid samples were in
saturated vapour pressure conditions (SVP); this was achieved by
measuring the vapour pressure of the samples on the top of the
sample cell by a Baratron pressure transducer. The liquid samples
were contained in a square flat aluminium cell (6 cm width, 6 cm
height, 0.5 cm thickness) placed in a liquid helium flow cryostat;
the sample temperatures were recorded by two Ge resistance
thermometers located at the upper and lower ends of the sample
cell, resulting in an average temperature $T= 1.96 K\pm \ 0.01 K$
throughout the measurements.  Experimental values of vapour
pressure for each mixture composition $x$ were found to be in
agreement with SVP data in the literature \cite{sydoriak}. For
each composition DINS spectra were recorded for runs lasting about
24 hours. The time-of-flight spectra were and normalized to the
monitor counts by using standard routines available on VESUVIO
\cite{evans}. In Figure 1 a typical time of flight spectrum as
recorded from a single detector for the $x=0.35$ mixture is
displayed.

 From this figure, one can note that the recoil peaks from the two
different atomic masses occur at well separated positions in the
time of flight spectrum. This is the case for the whole set of
data in the angular range explored.
 Due to the high values of wavevector
transfer accessed, the recoil peaks can be analyzed in wavevector
spaces, i. e. $y_3=\frac{M_3}{\hbar q}\left( \omega -\frac{%
\hbar q^{2}}{2M_3}\right) $ and $y_4=\frac{M_4}{\hbar q}\left( \omega -\frac{%
\hbar q^{2}}{2M_4}\right) $ for $^3$He and $^4$He respectively.
 The fixed-angle experimental
resolution, $R_{l}(y_{3,4})$, where $l$ is the $l-th$ fixed-angle detector
element, was determined for each detector through a standard VESUVIO
experimental calibration using a lead sample. The $R_{l}(y_{3,4})$ as in
previous measurements on $^{3}$He
 \cite{senesi,jltp} and $^{4}$He \cite{elioepl}, is well described
 by a Voigt function, whose parameters are $\sigma(y_3)=0.847$ $ \AA^{-1}$, $\frac{\Gamma}{2}(y_3)=1.371$ $ \AA^{-1}$ and
 $\sigma(y_4)=0.839$ $ \AA^{-1}$, $\frac{\Gamma}{2}(y_4)=1.740$ $ \AA^{-1}$, where $\sigma(y)$ is the standard deviation
 of the gaussian component and $\frac{\Gamma}{2} (y)$ is the half width at half maximum of the lorentzian component.
  A parallel procedure has also been set-up using Monte
 Carlo neutron transport simulation codes for the VESUVIO spectrometer \cite{ddnim,jltp,multiplo}
 in order to simulate the complete moderator-sample-detector neutron transport,
 including multiple neutron scattering and energy dependent neutron absorption. This procedure provided  simulated
 DINS measurements, a simulated resolution function $R_{l}(y),$ which  agreed with
both experimental data and  experimentally calibrated $R_{l} (y)$
\cite{ddnim,jltp,multiplo}. This ensured the reliability of the current
calibration procedure, and also allows the observed $J_{3,4}(y)$ to be
described by
   a convolution of the longitudinal momentum distribution and the instrumental
resolution function \cite{ddnim}.

\subsection{Absorption and multiple scattering correction}

The effects of the $^3$He neutron absorption on the measured
scattering from the liquid mixtures have been examined in detail,
using both an analytical approach as well as a deterministic Monte
Carlo simulation procedure. In $^3$He the neutron absorption cross
section is energy-dependent with the typical 1/v variation, with a
value of 5333 b for 25 meV neutrons \cite{mughabghab}. The
incident energy range covered by the measured $^4$He and $^3$He
recoil peaks were  9-15 eV and 13-20 eV respectively (using Au
absorption filters) . Although the absorption cross section of
$^3$He has a relatively smooth variation in these ranges, we
analyzed in detail the effects on the measured neutron Compton
profiles. An approach for the analytic correction of absorption in
strongly absorbing media was first proposed by Sears
\cite{sears1}. The double-differential scattering cross section is
calculated by evaluating the distribution of scattered neutrons
from a sample of finite size, with the quantities $S(q,\omega)$,
the scattering function, and $\Sigma(k)$, the total cross section
per unit volume for a neutron with wavevector k, occurring as
parameters of the Boltzmann equation in the neutron transport
theory \cite{sears2}.  In the $^3$He case
\begin{equation}
\Sigma(k)= \Sigma_a(k)+ \Sigma_s(k)\simeq \Sigma_a(k)\gg \Sigma_s(k)
\end{equation}
 where $\Sigma_a(k)$ and $\Sigma_s(k)$ are the absorption and
scattering cross sections per unit volume; the beam attenuation
due to multiple scattering is also found to  be negligible, since
the ratio of double to single scattering is of the order of
$\Sigma_s(k)/2\Sigma(k)$ \cite{sears2} . In the case of a slab
shaped sample and backscattering geometry, as shown in Figure 2,
the double-differential cross section for single scattering is
given by:
\begin{equation}
\frac{d^2\sigma}{d\Omega dE_f}=[\frac{A}{\Sigma(k_i)sec(\Phi_i)+
\Sigma(k_f)sec(\Phi_f)}] n_3 \sigma  \frac{k_f}{4\pi k_i}S(q,\omega)
\end{equation}\label{abs}
where  A is the surface area of the sample, $n_3$ is the $^3$He atomic number
density and $\sigma$ is the atomic scattering cross section. The wavevector
dependence of $\Sigma_a(k)$ is expressed by $\Sigma_a(k)=n_3 \sigma_a(k)= n_3
\frac{4\pi}{k} b_c" $ where $b_c"$   is the imaginary part of the coherent
scattering length\cite{sears2,sears3}.

Experimental time of flight spectra from different fixed-angle detectors have
been transformed into $y$ space and described in terms of the fixed-angle
neutron Compton profiles $J_l(y)$ using standard VESUVIO procedures; details
of the transformation of time of flight spectra in fixed-angle neutron
Compton profiles are accurately described in refs. \cite{evans,baciocco}; in
order to correct for the incident-wavevector dependent absorption an
analytical correction has been applied to the y-transformed data:
\begin{equation}
J(y)_{l,corr} = J(y)_l\frac{x}{x \sigma_3+(1-x) \sigma_4}[\sigma_a
(k_f)+\sigma_a(k_i) \frac{k^2_i}{k^2_f \cos\Phi_f}]\label{corr}
\label{cfact}\end{equation} where $\sigma_a(k)$ is the wavevector dependent
absorption of $^3$He, and $\sigma_3$ and $\sigma_4$ are the scattering cross
section for $^3$He and $^4$He respectively. In the case of inverse-geometry
spectrometers such as VESUVIO,
 the final wavevector is constant, and for a fixed angle spectrum, Eq. \ref{cfact}
assumes the form
\begin{equation}
J(y)_{l,corr} = J_l(y)[A + B \: k_i]
\end{equation}
where $A$ and $B$ are constants depending on the scattering angle and the
concentration $x$; finally, due the zero-order sum rule for J(y),
$J(y)_{l,corr}$ (the suffix $corr$ is then omitted in the next sections) have
been normalised to unity \cite{sears1}.
 The concentration dependent factor was first proposed by Hilton et al.
\cite{hilton} and is particularly valid under the current experimental
conditions of large wavevector and energy transfers, where cross correlation
between $^3$He and $^4$He cross sections are negligible.  As a complementary
procedure the Monte Carlo neutron transport code DINSMS \cite{multiplo} has
been also employed to evaluate multiple scattering contributions and to test
the analytical absorption corrections in the mixtures. In particular this
code accounts for the energy dependent absorption of the mixtures employing
the following expression:  :
\begin{equation}
n[x\sigma_3+ (1-x)\sigma_4]e^{-n[x(\sigma_3+\sigma_a)+ (1-x)\sigma_4]t}dt
\end{equation}
which represents the probability that a neutron travelling along
the $\widehat{t}$ direction within the slab-shaped sample will
scatter between $t$ and $t+dt$.
 As expected, the ratio of double to single scattering intensities varied
between $1.2\%$ for the $x=0.1$ mixture and $0.1 \%$ for the
$x=0.9$ mixture, in agreement with theoretical predictions
\cite{sears1,sears2}. Therefore multiple scattering corrections
were neglected.  Moreover, by comparing the ratio of simulated
data in momentum space with and without absorption contributions
included, the analytical correction factor, (Eq. \ref{corr}) was
recovered. An example of a correction factor, as derived from
simulations for a $x=0.35$ mixture for a scattering angle of
$135^{o}$ is presented in Figure 3.

\section{Data analysis and results}

The experimental spectra for each mixture composition were converted to
 $^3$He and $^4$He momentum space. Following this procedure a single function,
$J(y)$, averaged over all the 32 detectors was derived for each
isotope and for each composition.

The NCP were analysed by simultaneously fitting the two recoil peaks
appearing in $J(y_3)$ ($J(y_4)$). The component centered at $y=0$ in $J(y)$
was fitted by a model
 function $M(y)$ broadened by the instrumental resolution. The other component, centered at
 negative($^4$He) or positive ($^3$He)
 $y$ values were fitted
 using a Voigt function.
The model function, $M(y)$, used to describe the longitudinal momentum
distribution was of the form:

\begin{equation}
M(y)=
\frac{e^{\frac{-y^2}{2\sigma^{2}}}}{\sqrt{2\pi\sigma^{2}}}[\sum_{n=2l}^{\infty}d_n
H_n(\frac{y}{\sqrt{2\sigma^2}})]
\end{equation}

where $H_n(\frac{y}{\sqrt{2\sigma^2}})$ is the \emph{n-th} Hermite
polynomial, and $\sigma$ and $d_n$ are fitting parameters; this
functional form was applied by Sears \cite{searspr69} for the
analysis of neutron scattering from pure liquid $^4$He, and a
generalized form, including angular dependencies, is currently
used for momentum distribution spectroscopy in hydrogen containing
systems on the VESUVIO spectrometer \cite{reiterprl}. We used
$d_0=1$ and $d_2=0$, in order that $M(y)$ satisfies the following
sum rules:

\begin{equation}
\int_{-\infty}^{\infty} M(y) \; dy=1
\end{equation}

\begin{equation}
\int_{-\infty}^{\infty} M(y)\; y^2 \; dy=\sigma^2=\sigma_y^2
\end{equation}

In the present case orders up to $H_4(\frac{y}{\sqrt{2\sigma^2}})$
were employed. The inclusion of higher order polynomials did not
result in significant improvements of the fits. As an example, in
Figure 4 the $J(y_3)$ function for three different compositions is
presented, together with the fitted lineshapes. The results for
the determination of the mean kinetic energies for the two
isotopes and the six mixture compositions are reported in Table 1,
and are shown in Figure 5, in comparison with previous experiments
and several theoretical predictions.

The present results  extend the range of concentration x with
respect to previous DINS measurements, with improved statistical
accuracy. The $^4$He kinetic energy is found to decrease with
concentration, while the $^3$He kinetic energy does not depend
appreciably on the concentration.

\section{Discussion and conclusions}

The results of the present experiment show that the kinetic energy
of $^4$He atoms is strongly affected by the addition of $^3$He;
this is expected from density and quantum statistics arguments. A
remarkable agreement with previous measurements and recent
theories is found, with slight discrepancies with respect to
finite temperature method calculations \cite{boninsegni}. On the
other hand the kinetic energy of the $^3$He atoms appears
unaffected by the presence of the higher density boson fluid,
which seems to promote $^3$He delocalization. It is illustrative
to compare the $J(y_3)$ spectra for the $x=0.10$ mixture and for
the pure $^3$He fluid, as shown in Figure 6, there is no
substantial broadening of $J(y_3)$ on going from the pure liquid
to the low-concentration mixture.

We stress that in several papers \cite{bor97,mazzanti,boninsegni}
the discrepancies between experiments and theories on the
determinations of $\langle T \rangle_3 $ have been attributed to
high energy tails in the dynamic structure factors, resulting in
exponential-like tails in the momentum distributions, which could
be masked by  background noise, or are not accounted for using
inadequate model functions for $J(y_3)$
\cite{bor97,mazzanti,boninsegni}. We point out, however, that
since the high energy tails are due to the repulsive part of the
interatomic potential, they are present in the momentum
distribution of the $^4$He component as well. This is well
illustrated by Fig. 6 of ref. \cite{bor97}, where the momentum
distributions of $^3$He and $^4$He  in an $x=0.066$ mixture have
practically the same high-momentum tails. The background noise is
the same for the two isotopes in the same measurement, while the
spectrometer resolution narrows on going from variables $y_4$ to
$y_3$ in the same spectra; therefore one would expect an increased
sensitivity to the detection of the high-momentum tails in
$J(y_3)$.  To test this hypothesis we simulated a measurement of
$J(y_3)$ for an $x=0.066$ mixture in the experimental
configuration of the present measurements. The momentum
distribution published in ref. \cite{bor97}, $n^{(3)}(k)$ of
$^3$He for the $x=0.066$ mixture at zero temperature was converted
to $J(y_3)$ using the general result \cite{momentum}:
\begin{equation}
J(y)= \frac{1}{2 \pi}\int_{|y|}^\infty dk\; k\; n(k)
\end{equation}
This function shows pronounced high-momentum contributions due to
the depletion of the Fermi sphere, the Fermi wavevector being
$k_f=0.347 \AA^{-1}$ . The kinetic energy value, obtained from the
second moment of the reconstructed $J(y_3)$ was  $\langle T
\rangle_3= 18.2 K $. The function was used as input for the DINSMS
code, which was run for a scattering angle of $135^{o}$. The input
$J(y_3)$ and the simulated experiment are reported in Figure 7.

The simulated spectrum was fitted by the convolution of a Voigt
function, representing the spectrometer resolution, and two model
functions : a simple gaussian and the Gauss-Hermite expansion
introduced above (Section III) with orders up to
$H_8(\frac{y}{\sqrt{2\sigma^2}})$. The use of a single gaussian
resulted in $\langle T \rangle_3= 16.7 \pm 0.4 K $, while the use
of the Gauss-Hermite expansion resulted in $\langle T \rangle_3=
17.0 \pm 0.4 K $. This indicates that the experimentally
determined kinetic energies of $^3$He are altered by less than 2.5
K, ruling out a strong effect of the high-momentum tails on the
determination of kinetic energies.

A second test was performed to compare self-consistently the
datasets available from the present measurements. The pure $^3$He
liquid data ($x=1.00$) were used as the calibration measurement,
and $J(y_3)$ was modelled to have the functional form reported in
ref. \cite{azjltp}, corresponding to a Fermi-like momentum
distribution with a discontinuity at the Fermi momentum and
high-momentum exponential tails; the resulting kinetic energy was
$\langle T \rangle_3= 11.7 K $, a value close to the theoretical
predictions \cite{casulleras}. This function was
 convoluted with a Voigt function, an "effective" resolution to be determined
 from the fit to the pure liquid data. The resulting Voigt function had the
 following parameters:$\sigma(y_3)=1.262$ $ \AA^{-1}$, $\frac{\Gamma}{2}(y_3)=0.9173$ $
 \AA^{-1}$. This "effective" resolution was employed to fit the $x=0.1$ data,
 and the resulting kinetic energy was $\langle T \rangle_3= 10.47 \pm 0.6 K
 $, confirming that no increase in kinetic energy, even in low-concentration mixtures,
 is indicated in the DINS experiments.

 This picture suggests that the local environment
of the $^3$He atoms remains unchanged in saturated vapour pressure liquid
mixtures. Within the same picture, the $\langle T \rangle_3 (x) $ behaviour
results also in a $\langle T \rangle_3 (n) $ behaviour, where $n$ is the
total atomic density, which differs radically from the widely assessed
atomic-density dependence of kinetic energy of all quantum fluids and solids.
Figure 8 shows the $\langle T \rangle_3 (n) $ behaviour of the mixtures, as
compared to $\langle T \rangle_3 (n) $ of the pure liquid
\cite{senesi,dimeo}. The atomic density in the mixture is higher than in pure
liquid $^3$He in equilibrium, and increases up to $n = 21.87 nm^{-3}$ for $x
= 0$, upon adding $^4$He \cite{kerstead}. A statistically significant
departure from the density-dependence observed in the pure fluid appears for
$n \geq$ 19 nm$^{-3}$. However we stress that in this case the simultaneous
changes of density and concentration prevent a thorough picture of the
density and concentration dependence of $\langle T \rangle_3 $. Systematic
studies in several (concentration, pressure, temperature) thermodynamic
states not previously investigated both experimentally and theoretically are
certainly needed on the experimental and theoretical sides. For example,
$\langle T \rangle_3 (n) $ measurements on mixtures at fixed concentration
(fixed concentration and increased pressure and density) would test whether
the density dependence is recovered upon approaching and crossing the
liquid-solid transition. However previous and present results show
unambiguously a behaviour of $\langle T \rangle_3 $ in the mixture which is
not density dependent as for other quantum fluids and is not reproduced by
any simulation studies.

The interpretation of these results is far from obvious. Two independent
measurements \cite{wang, az3he4he} and the present work, have shown
unequivocally that the $^3$He kinetic energy in the mixtures is essentially
independent on concentration and density. Further, we have shown that high
energy tails cannot explain this surprising result.  Given the fact that a
remarkable agreement between experiment and theory has been found for the
pure helium liquids, we hope that our work stimulates further theoretical and
experimental effort. From the experimental point of view, we can envisage
that high-resolution and high statistical quality data over an extended range
of temperatures, compositions and densities, can be obtained employing novel
chopper and resonance filter spectrometers at pulsed neutron sources.
However, the understanding of these results remains a challenge to
conventional theories of liquid isotopic helium mixtures.

\begin{acknowledgments}

This research has been supported by the VESUVIO project (EU
Contract Number: ERBFMGECT980142). RS kindly acknowledges J.
Boronat, F. Mazzanti and A. Polls for stimulating discussions. The
authors acknowledge the ISIS User Support Group for the valuable
technical support during experimental measurements.
\end{acknowledgments}

\pagebreak
\section{tables}

\begin{table}
\caption{Single-particle kinetic energies for $^3$He and $^4$He respectively
for the six mixtures. The density of the mixtures is also reported; density
values were derived extrapolating the data from Table VI of ref.
\cite{kerstead}to $T= 1.96 K$. Note that due to the very small intensities of
the $^4$He recoil peaks for the $x=0.65$ and the $x=0.90$ mixtures it was not
possible to reliably determine $\langle T \rangle_4 $.}
\begin{tabular}{|c|c|c|c|}
\hline x & n (nm$^{-3}$) & $\langle T \rangle_3 $ (K) & $\langle T \rangle_4 $ (K)  \\
\hline\hline 0.00 & 21.87 &- & 16.0$\pm $0.5 \\
\hline 0.10 & 21.4 & 12.1$\pm $0.4 & 13.8$\pm $0.6  \\
\hline 0.35 & 19.94 & 10.4$\pm $0.3 & 12.0$\pm $0.6  \\
\hline 0.65 & 18.22 & 11.8$\pm $0.7 & - \\
\hline 0.90 & 16.27 & 10.7$\pm $0.8 & - \\
\hline 1.00 & 15.44 & 10.9$\pm $0.4 & - \\
\hline
\end{tabular}
\end{table}

\section{Figure captions}

Figure 1. Bottom: time of flight DINS spectrum from the $x=0.35$
mixture
 for a single detector at the scattering angle $2 \theta
=135 ^{o}$. The $^3$He and $^4$He signals occur at approximately
205 $\mu s$ and 240 $\mu s$ respectively; the sample cell signal
is located at about 340 $\mu s$. Top:
 wavevector and energy transfer range accessed in the bottom spectrum.

Figure 2. Schematic diagram of the scattering geometry for
analytical absorption corrections. S is the slab-shaped absorbing
sample. $\Phi_{i,f}$ and $k_{i,f}$ are the angle of incidence
(scattering) with respect to the normal to the sample, and initial
(final) neutron wavevectors. Note that for the present case
$\Phi_i =0$.

Figure 3. Correction factor derived from the ratio of simulated
data with and without absorption contributions.

Figure 4. $J(y_3)$ for the $x=0.10, 0.35,$ and $1.00$ mixtures
(solid circles); The peak centered at negative values of momentum
 corresponds to the $^4$He peak, centered at lower recoil energy.
 The solid lines are the fits to the neutron Compton profiles.

Figure 5. Single-particle kinetic energies for $^3$He(lower panel,
solid circles) and $^4$He (upper panel, solid circles)
respectively for the six mixtures. DINS experimental results from
Azuah {\it et al.} \cite{az3he4he}, right-triangles. DINS
experimental results from Wang {\it et al.} \cite{wang},
left-triangles. Variational calculations at zero temperature for
x=0.066 from Boronat {\it et al.} \cite{bor97}, solid squares.
Restricted path integral Monte Carlo calculations at $T= 2 K$ from
Boninsegni {\it et al.} \cite{boninsegni}, open diamonds.
Variational Monte Carlo calculations at zero temperature from Lee
{\it et al.} \cite{lee}, solid triangles. Restricted path integral
Monte Carlo calculations at $T= 2 K$ from Boninsegni {\it et
al.}\cite{ceperleymix}, solid diamond.

Figure 6. $J(y_3)$ for the $x=0.10$ mixture, full circles and pure
$^3$He liquid ($x=1.00$), open triangles.

Figure 7. Simulated $J(y_3)$ for a $x=0.066$ mixture. Upper panel,
input Neutron Compton Profile derived from ref. \cite{bor97};
lower panel, the resulting $J(y_3)$ after the simulation with the
DINSMS code.

Figure 8. $\langle T \rangle_3 (n) $ for $^3$He-$^4$He liquid
mixtures and pure $^3$He liquid; present measurements on the
mixtures, solid circles; DINS experimental results from Azuah {\it
et al.} \cite{az3he4he}, on the mixtures, right-triangles; pure
liquid $^3$He from Dimeo {\it et al.} \cite{dimeo},open squares;
high density pure liquid $^3$He from Senesi {\it et al.}
\cite{senesi},open circle.

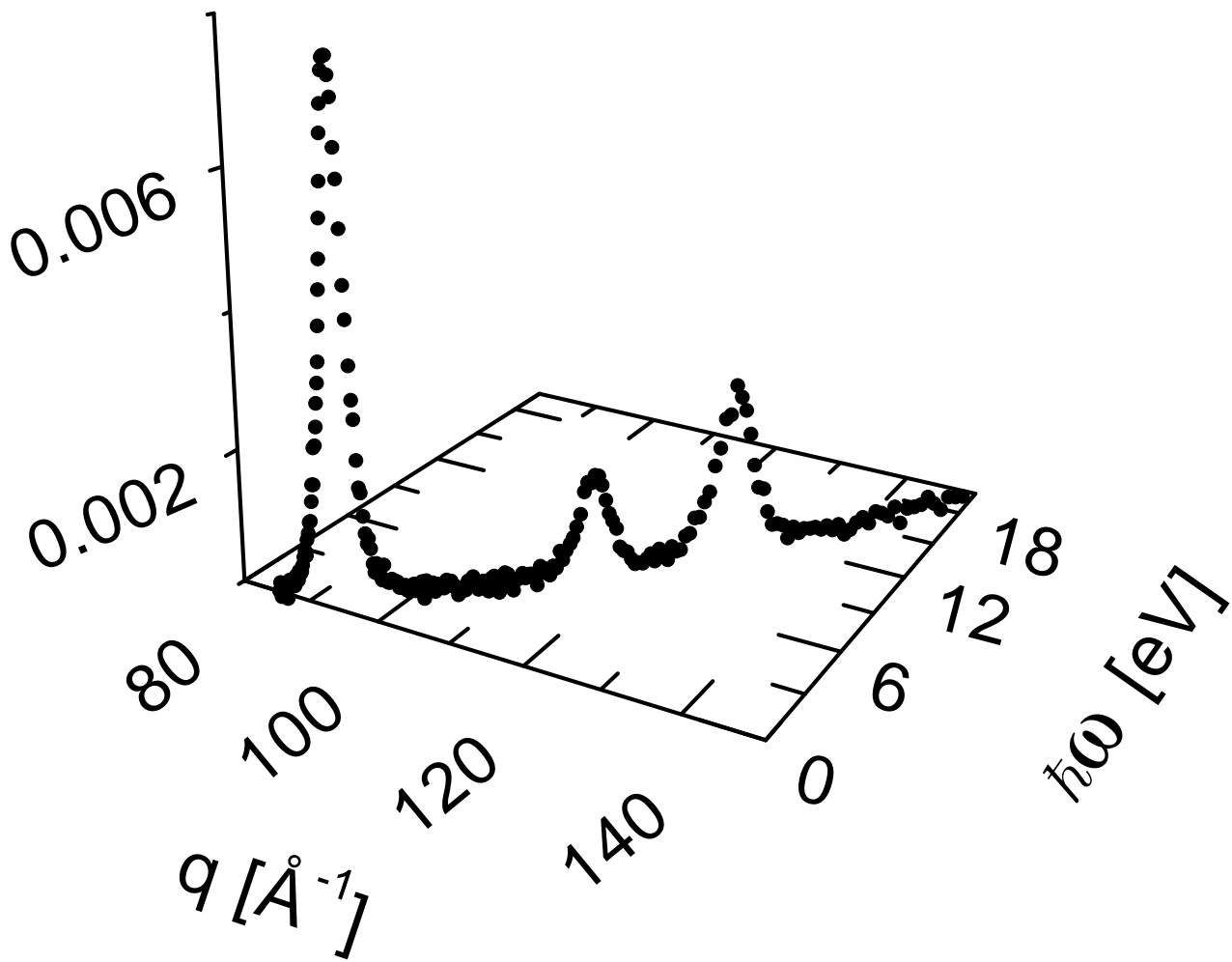
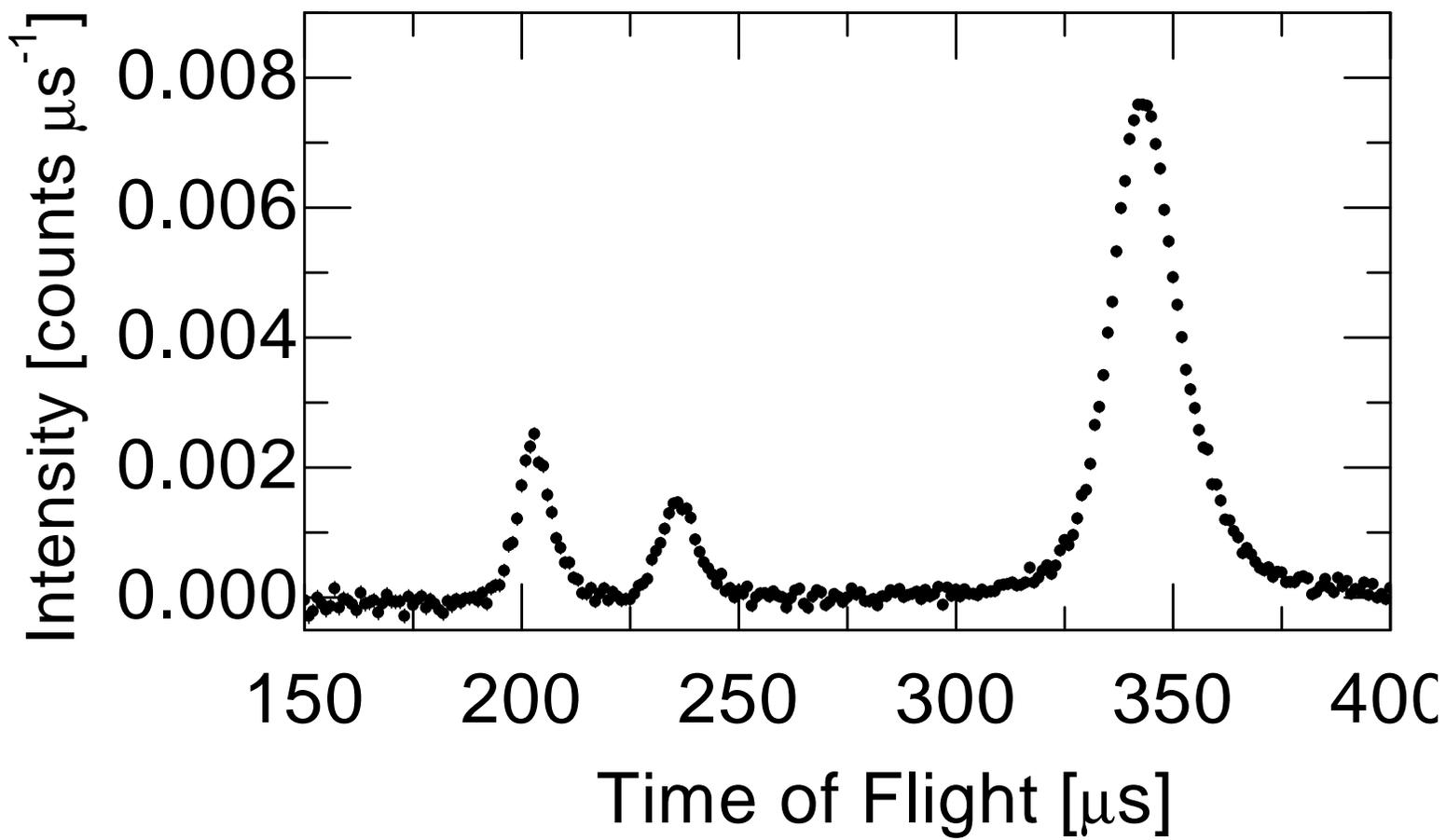
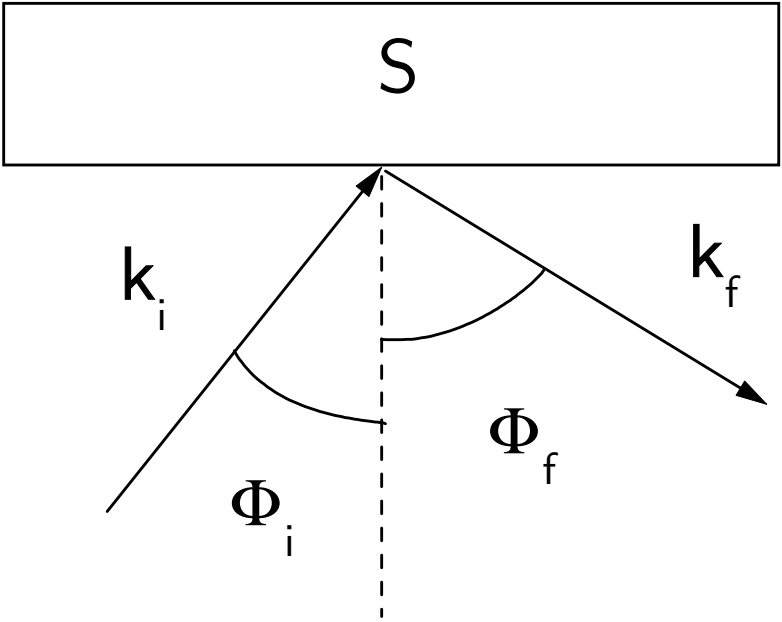
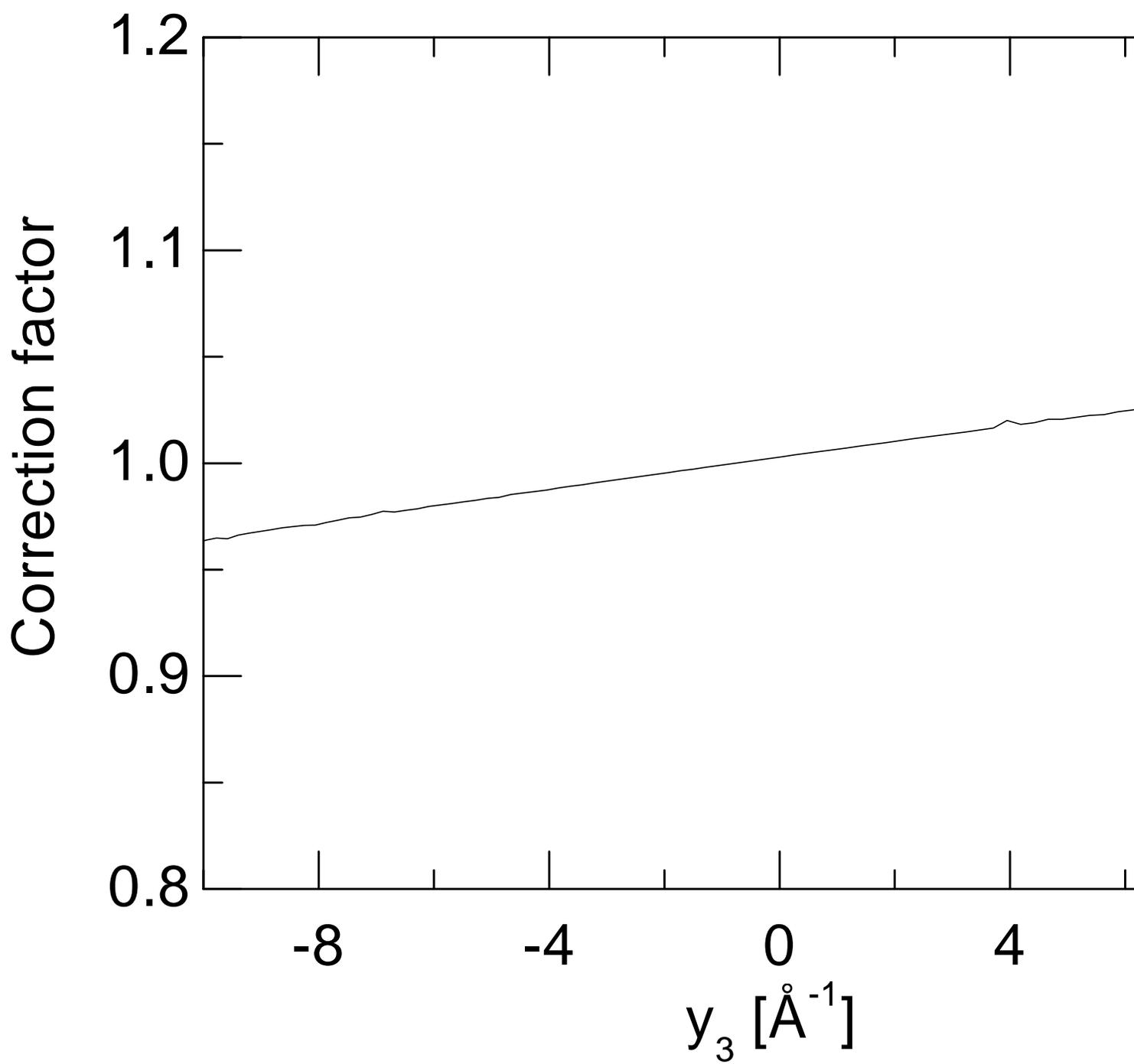
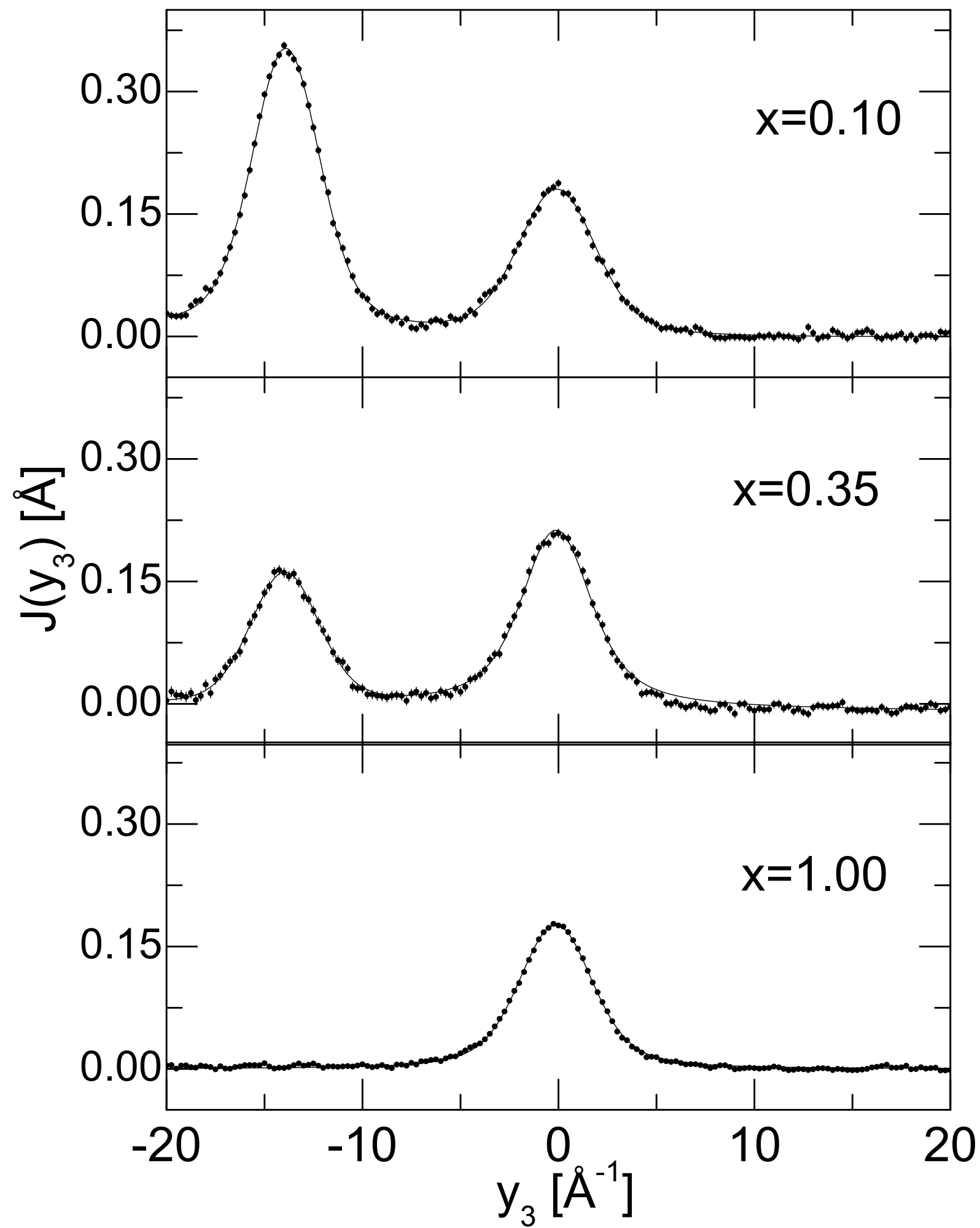
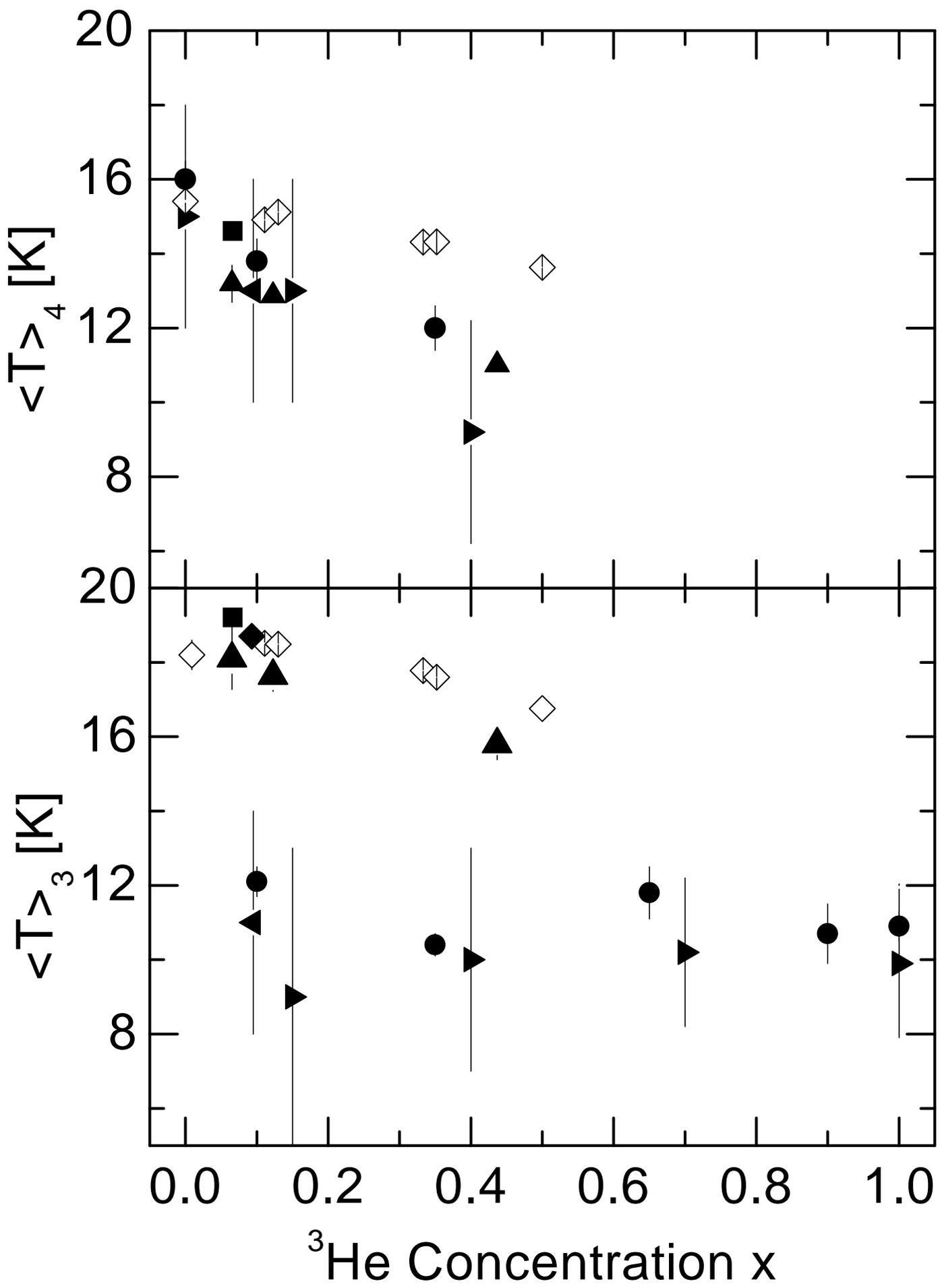
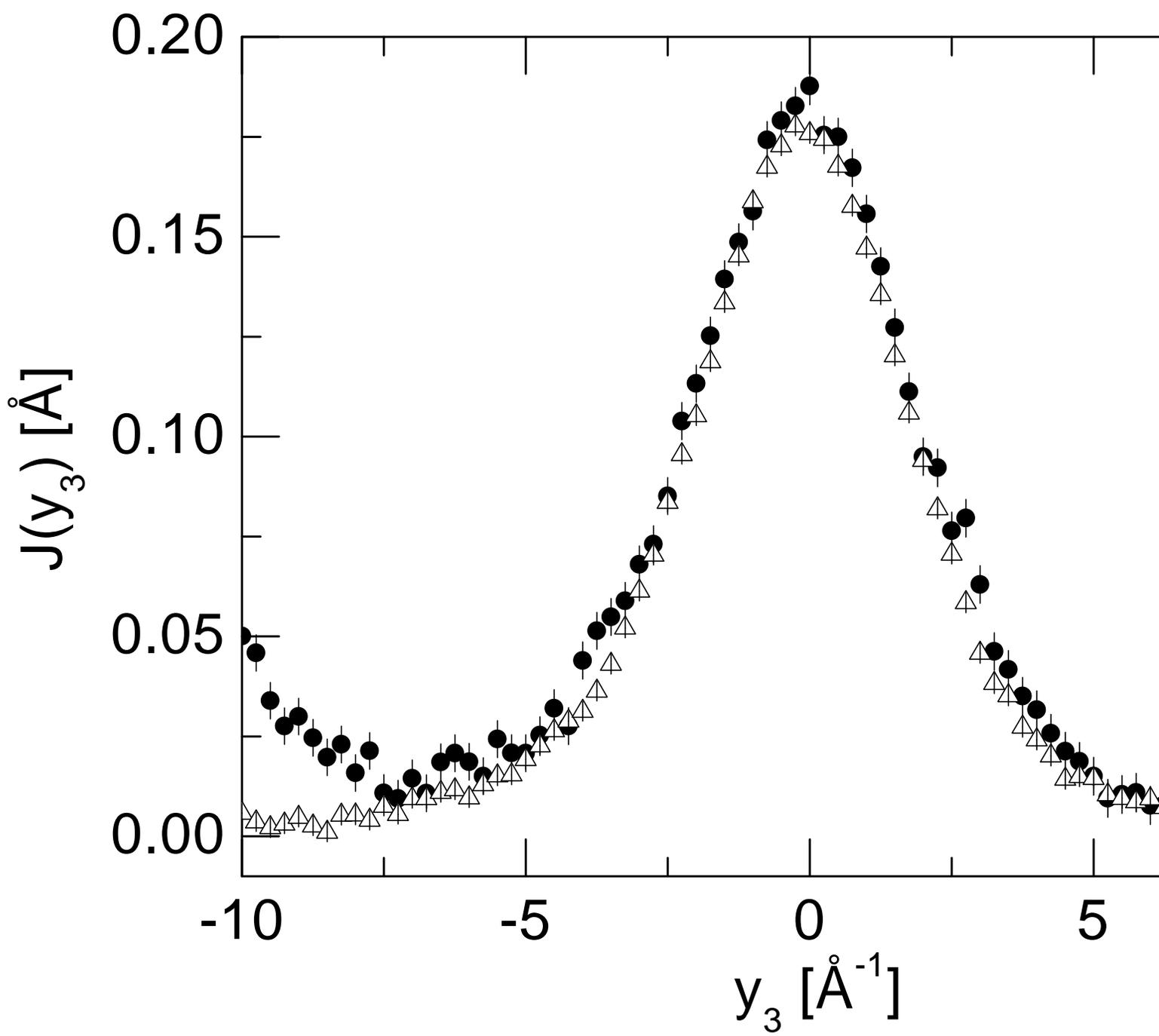
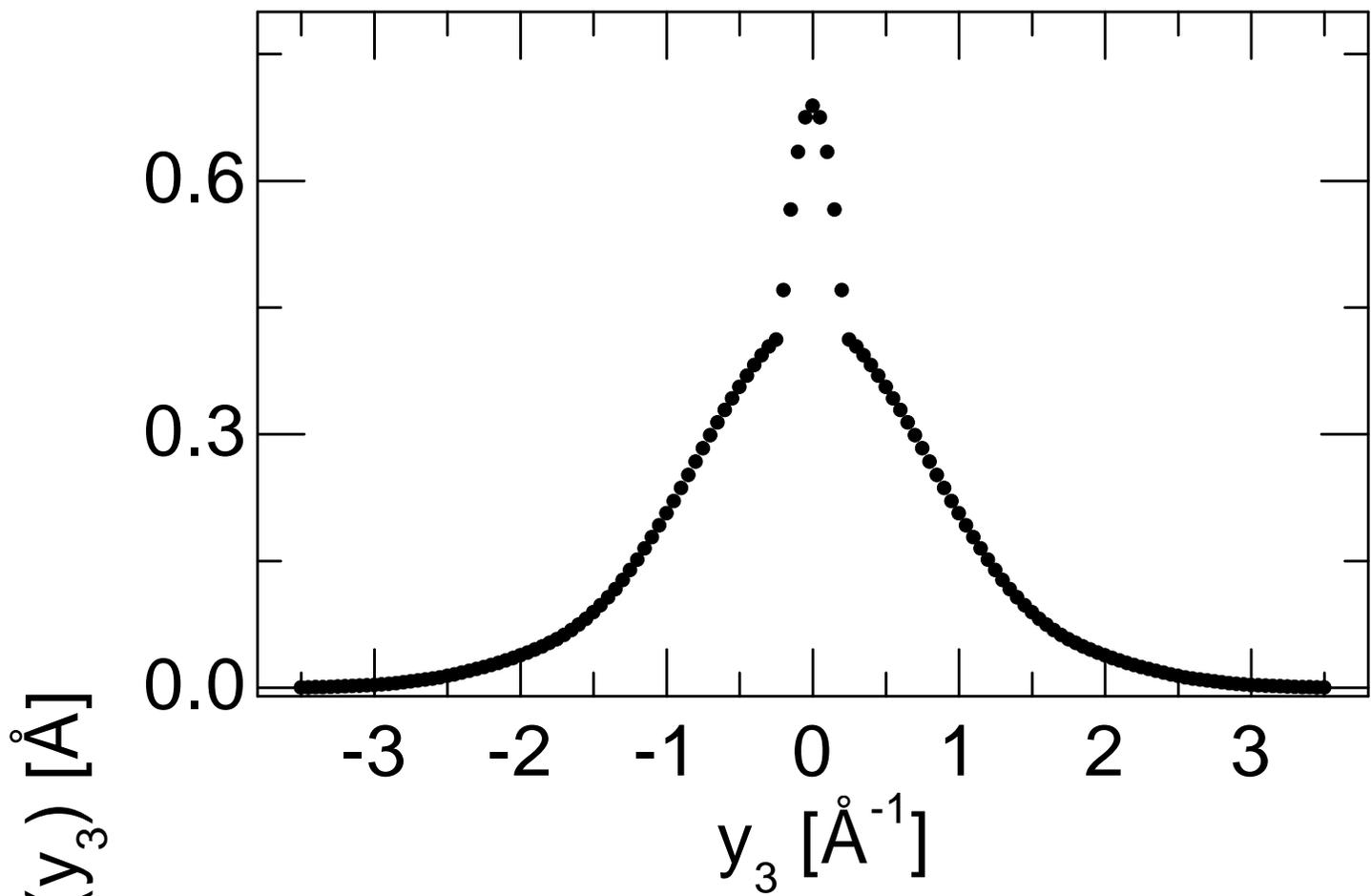
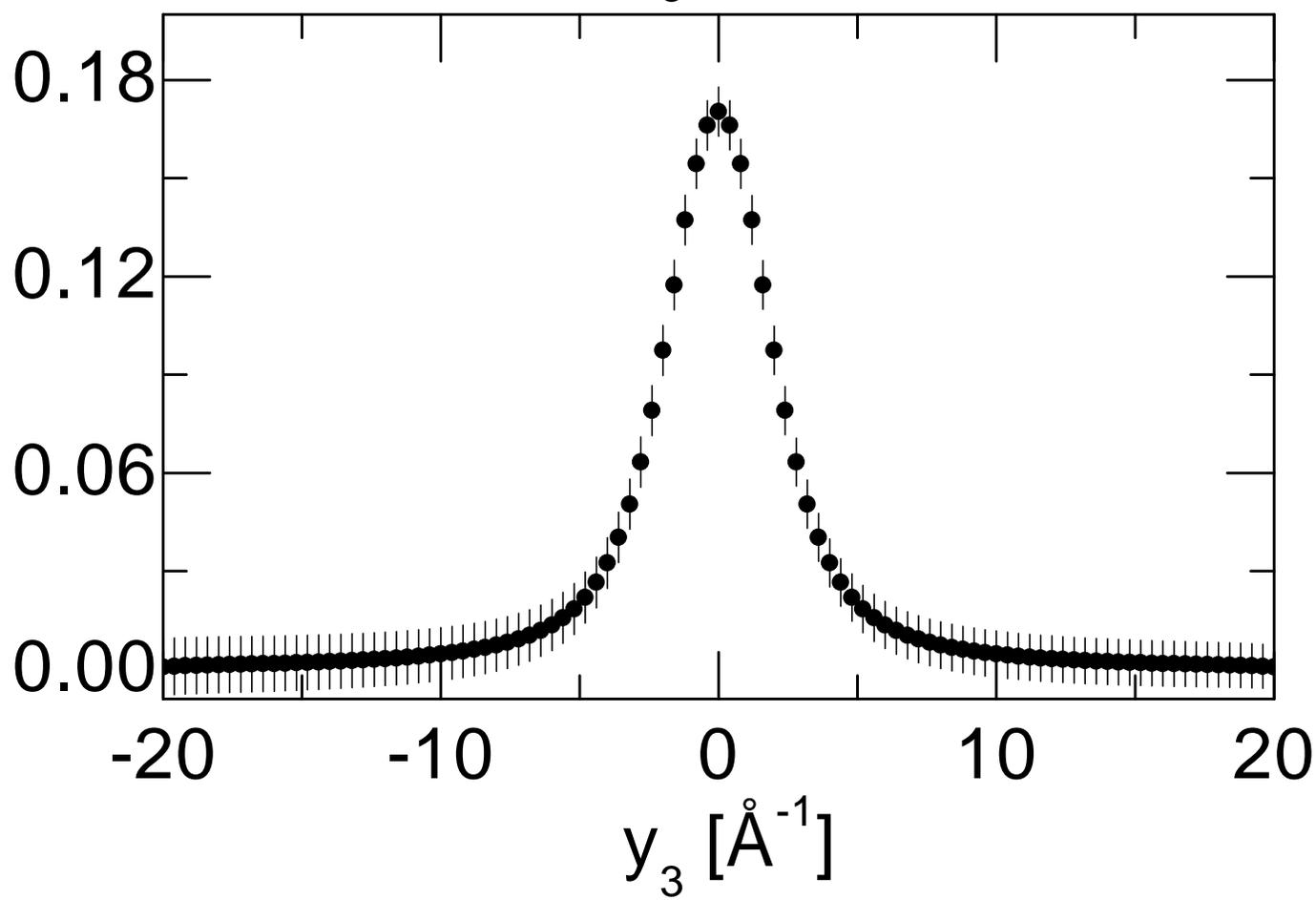
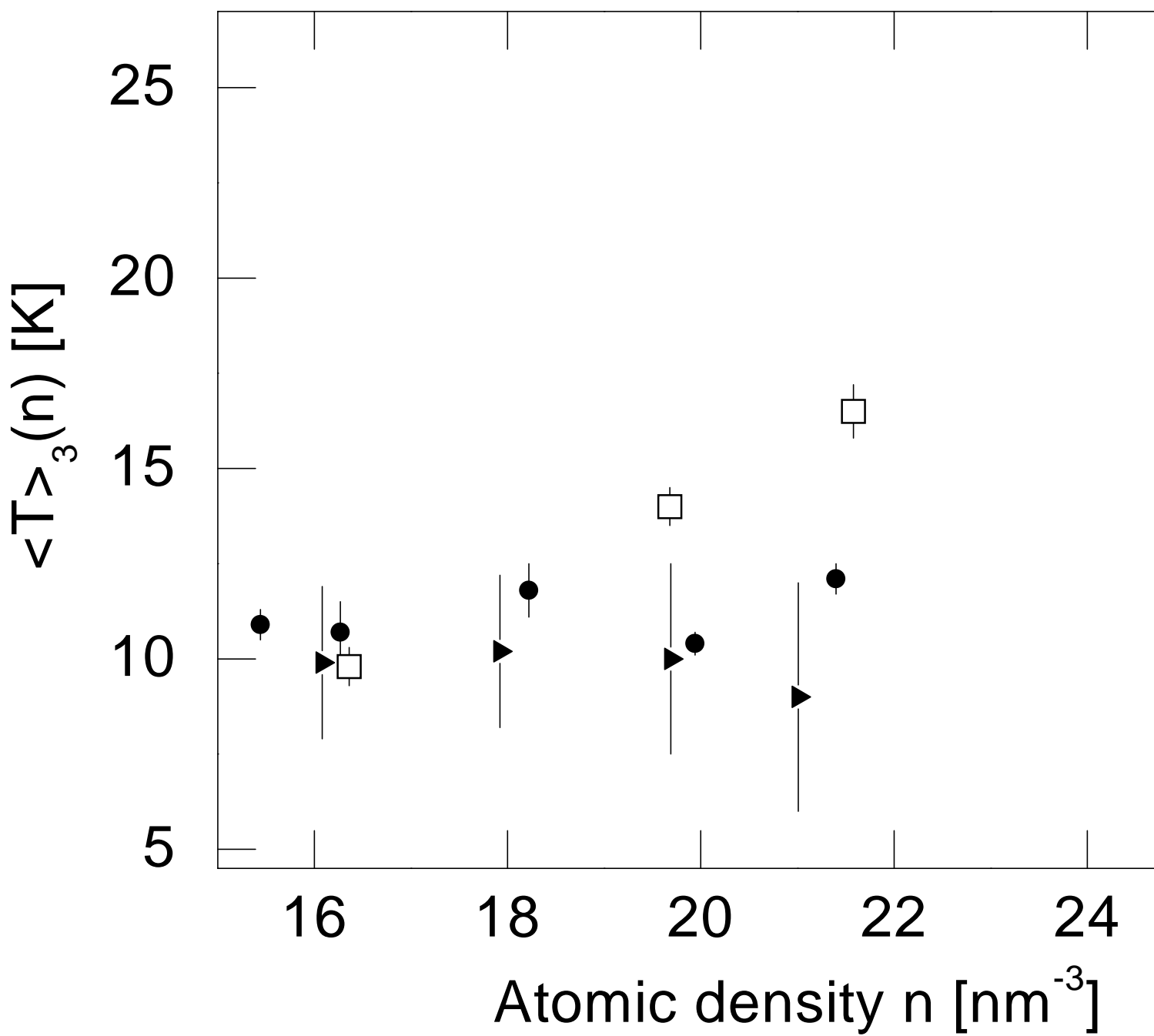
\end{document}